\documentclass[pdftex,twocolumn,epjc3]{svjour3}          

\RequirePackage[T1]{fontenc}

\smartqed  

\RequirePackage{graphicx}
\RequirePackage{mathptmx}      
\RequirePackage{flushend}
\RequirePackage[numbers,sort&compress]{natbib}
\RequirePackage[colorlinks,citecolor=blue,urlcolor=blue,linkcolor=blue]{hyperref}

\journalname{Eur. Phys. J. C}

\usepackage{amsmath}
\usepackage{amsfonts}
\usepackage{amssymb}%

\begin{document}
\title{Repulsive Casimir force in stationary axisymmetric spacetimes}

\author{V. De La Hoz-Coronell \thanksref{e1,addr1}
        \and
        A.E. Gon\c{c}alves \thanksref{e2,addr1}
        \and
        M.C. Baldiotti \thanksref{e3,addr1}
        \and
        R.C. Batista \thanksref{e4,addr2}
}

\thankstext{e1}{e-mail: vdcoronell@uel.br}
\thankstext{e2}{e-mail: goncalve@uel.br}
\thankstext{e3}{e-mail: baldiotti@uel.br}
\thankstext{e4}{e-mail: rbatista@ect.ufrn.br}

\institute{Departamento de F\'isica, Universidade Estadual de Londrina, 86051-990,
Londrina, PR, Brazil\label{addr1}
          \and
          Escola de Ci\^{e}ncias e Tecnologia, Universidade Federal do Rio Grande do
Norte, 59072-970, Natal, RN, Brazil\label{addr2}
}

\date{Received: date / Accepted: date}
\maketitle

\begin{abstract}
We study the influence of stationary axisymmetric spacetimes on Casimir
energy. We consider a massive scalar field and analyze its dependence on the
apparatus orientation with respect to the dragging direction associated with
such spaces. We show that, for an apparatus orientation not considered before
in the literature, the Casimir energy can change its sign, producing a
repulsive force. As applications, we analyze two specific metrics: one
associated with a linear motion of a cylinder and a circular equatorial motion
around a gravitational source described by Kerr geometry.

\end{abstract}
\maketitle











\section{Introduction}

In its original form, the Casimir effect is a quantum phenomenon that arises
in the vacuum state of the electromagnetic field in the presence of two
neutral metal parallel plates, which impose boundary conditions to the field
and produce attraction force between the plates \cite{Casimir1, Casimir2}. A
problem that have attracted a considerable attention in recent years is on
which conditions the Casimir force can change from attractive to repulsive,
see \cite{JeanWi2018} and references therein.

The original attractive force results from a negative Casimir energy. We are
especially interested in conditions that can cause a change in the sign of the
Casimir energy. It can happen, for example, if a mixed boundary condition (BC)
is considered. That is, if Dirichlet or Neumann boundary condition is imposed
on both plates, this force is attractive, but it becomes repulsive when fixing
a plate with the Dirichlet boundary condition and the other plate with Neumann
boundary condition \cite{Mostepanenko}. In the context of dielectrics
materials, Lifshitz predicted in 1956 that the force is attractive for two
identical dielectric plates in vacuum \cite{Liftshitz2}. A few years later, in
1961, Lifshitz and colleagues generalized this result by considering a medium
between the dielectrics plates \cite{Liftshitz}. As a result, they showed
that, if the two plates and the medium have different dielectric constants,
the force can become repulsive. Experimental verification of the repulsive
Casimir effect was performed at \cite{Capasso}, by filling the medium with a
dielectric liquid, such that the dielectric constants of the three bodies
involved in the experiment differ causing a repulsive effect. However, if the
two plates have the same dielectric constant, the force is always attractive,
regardless the dielectric constant of the medium \cite{Liftshitz}. Indeed,
this last conclusion is a particular result of a non-go theorem \cite{Israel},
which states that \textquotedblleft the Casimir force between two dielectric
objects, related by reflection, is attractive.\textquotedblright\ A possible
\textquotedblleft loophole\textquotedblright\ in this theorem may arises when
a chiral material is considered as a medium between the plates
\cite{JeanWi2018}.

In the context of curved spacetimes, Ref. \cite{Saharian} showed that, in the
de Sitter spacetime, for a massive scalar field minimally or conformally
coupled to the curvature, the Casimir force can change its sign (for the same
BC) if the proper distances between the plates is larger than the curvature
radius. We stress that this effect is a consequence of a coupling between the
field and the spacetime curvature. Other important result is exposed in the
quantum cosmology landscape at \cite{Odintsov}, where it is shown that the
character of being attractive or repulsive Casimir effect is due to the choice
of boundary condition related with the dynamic properties of the scale factor
in an expanding Friedmann-Robertson-Walker universe.

Other works have considered the Casimir plates immersed in stationary
axisymmetric spacetime, with the assumption of the apparatus is very small
compared with typical scale on which the metric varies. In this case, the
Casimir effect does not break the equivalence principle \cite{Milton1,Milton2}%
. The influence of the gravity in the Casimir energy in such scenario was
studied for some specific geometries. For example, the Kerr spacetime in
\cite{Sorge}, which study the corrections of the Casimir energy due to the
influence of a rotating gravitational source, for a Casimir apparatus
describing a circular equatorial orbit. In the aforementioned work, the plates
of the apparatus are oriented parallel to the radial coordinate of the
gravitational source. The extension for a general stationary spacetime was
presented in \cite{zhang}.

From the above discussion we see that, in a stationary spacetime, the Casimir
force is known to change from attractive to repulsive in three cases: for a
mixed boundary conditions; with change of the media between the plates (chiral
material) and as a consequence of a non-null Ricci scalar curvature. In this
work we present a new case when this change can occur. Namely, the change in
the sign of the (static)\ Casimir energy of an identical plates apparatus, for
a massive scalar field described by a vacuum solution of a stationary
axisymmetric metric. In this case, the Ricci scalar curvature\ $R$ is zero and
the change of sign cannot be associated with direct coupling between the
scalar field and $R$. As we will show, in this scenario the Casimir force
changes sign when the plates are parallel to the direction of spacetime drag,
a case not considered before in the literature. This effect is related with
the presence of intrinsic non-diagonal terms in the metric, that is, metrics
in which the timelike Killing field fails to be globally hypersurface-orthogonal.

This paper is organized as follows, in Sec. \ref{kge} we consider a massive
scalar field immersed in a general axisymmetric spacetime, where we solve the
Klein-Gordon equation and determine the eigenfrequencies and the normalized
solutions. In the Sec. \ref{casimir} we compute the Casimir energy and present
our main result, i.e., the fact that Casimir energy change sign due to the
orientation of the apparatus with respect to the drag of the spacetime, we
also establish the connection with some already known results for massless
case. In the Sections \ref{cylinder} and \ref{kerr} we apply our approach for
two specific stationary background geometries: one with cylindrical symmetry
and Kerr metric, respectively. Some final remarks are presented in Sec.
\ref{final}.

Throughout this paper we use natural units $\hslash =c=G=1$ and we work in
signature metric equal to $-2$.

\section{\label{kge}Klein-Gordon equation and the frequencies}

We are interested in vacuum solutions for massive scalar field inside a
Casimir apparatus in an axially symmetric stationary spacetime. Following
Sloane and Chandrasekhar \cite{Sloane,chandrasekhar}, we demand that such
spacetime is stationary, $x^{0}\rightarrow x^{0}+c_{0}$, has axial symmetry,
$x^{1}\rightarrow x^{1}+c_{1}$ and is invariant under simultaneous reflection
with respect to both $x^{0}$ and $x^{1}$, $x^{0}\rightarrow-x^{0}$ and
$x^{1}\rightarrow-x^{1}$. Under these assumptions, the most general line
element is given by:
\begin{align}
ds^{2} &  =g_{00}\left(  dx^{0}\right)  ^{2}+2g_{01}dx^{0}dx^{1}+g_{11}\left(
dx^{1}\right)  ^{2}\nonumber\\
&  +g_{22}\left(  dx^{2}\right)  ^{2}+g_{33}\left(  dx^{3}\right)
^{2}\,,\label{st-axi-sym}%
\end{align}
where the components of the metric tensor depend only on coordinates $x^{2}$
and $x^{3}$. The above metric reflects the non-reversal of time. For a
stationary (non-static) spacetime, it is related to the lack of a global
time-oriented Killing field (the time-like Killing field is no longer
hypersurface orthogonal). A well-known example, which will be analyzed as a
special case of our development, is Kerr's geometry. In this case the cross
term, $g_{01}$, is associated with the rotation of the gravitational field
source. However, $g_{01}$ is not necessarily related with a rotating source
\cite{CanSc2003}. Nonrotating vacuum solutions with this term can be used,
e.g., to describe superconducting strings with linear momentum
\cite{GeliTi2000}.

Our goal is to analyze the impact of scalar field mass and the apparatus orientation with respect to the symmetry axis, $x^{1}$, on the Casimir energy. We also compare our results with previous ones, where a specific orientation was chosen for the massless field case \cite{Sorge,zhang}.
Since energy is a frame-dependent quantity, we must choose the same coordinate frame used in
these papers. Namely, a local Cartesian coordinate frame $\left(
x,y,z\right)$ comoving with the apparatus. In this frame we have the
following line element,
\begin{equation}
ds^{2}=g_{tt}dt^{2}+2g_{tx}dtdx+g_{xx}dx^{2}+g_{yy}dy^{2}+g_{zz}%
dz^{2}~,
\label{eq:LinEleCon}%
\end{equation}
with determinant $g$,
\begin{equation}
g=g_{yy}g_{zz}\tilde{g}~,\ \tilde{g}=g_{tt}g_{xx}-g_{tx}^{2}~.\label{eq:GTil}%
\end{equation}
A time-dependent transformation applied to the metric 
(\ref{eq:LinEleCon}) may change the notion of energy, making 
comparisons with the previous results meaningless.

Let us consider a scalar field $\psi\left(  x\right)$, with minimal coupling
to gravity and mass $m$, that obeys the Klein-Gordon (KG)\ equation, which in
a curved spacetime reads%
\begin{equation}
\hat{D}\psi=0~,\ \hat{D}\equiv\frac{1}{\sqrt{-g}}\partial_{\mu}\sqrt{-g}%
g^{\mu\nu}\partial_{\nu}+m^{2}~.
\end{equation}
We call $\hat{D}$ the KG operator. Now, we assume the same
approximation made in \cite{Sorge,zhang}. Namely, that the apparatus has a
small size compared to the scale on which the metric varies. In this situation, one can
consider a zero-order expansion of $g_{\mu\nu}$ around the origin. At
a first glance, the constant metric approximation on the Klein-Gordon equation
makes the problem appear equivalent to the flat case. However, it has been
demonstrated by Sorge \cite{Sorge} and Zhang \cite{zhang} that this situation
is not equivalent to the one in flat space-time. As pointed out in
\cite{Sorge}, this unexpected behavior is due to a symmetry breaking. Namely,
for a Casimir apparatus in an axial symmetric metric, with non-diagonal
element $g_{tx}$ and the plates orthogonal to the $x$-direction, the associated break of translational invariance induces a
distortion in the discretized field modes inside the cavity. In \cite{zhang}
it is shown that, even in this constant metric approximation, the breaking of
translational invariance induces non-trivial corrections in Casimir
thermodynamic description. For more developments and discussion in this
context see \cite{ValBeM2019,BezCuFM2017}. Our goal is to show that, not only
this translational invariance, but also the apparatus orientation with respect
to the $x$-direction, influences the discretized field modes inside
the cavity. In this zero order expansion of the metric, the KG operator takes
the form
\begin{equation}
\hat{D}=g^{tt}\partial_{t}^{2}+2g^{tx}\partial_{t}\partial_{x}+g^{xx}%
\partial_{x}^{2}+g^{yy}\partial_{y}^{2}+g^{zz}\partial_{z}^{2}+m^{2}~.
\end{equation}
For constants $K,~N~,\omega$, and considering a solution of the form%
\begin{equation}
\psi=Ne^{-i\omega_{n}t}e^{ik_{z}z}\exp\left(  iK\omega x\right)  f\left(
x\right)  g\left(  y\right)  ~,\label{eq:SolKG}%
\end{equation}
we have%
\begin{align}
\hat{D} &  =\left(  2g^{tx}K-g^{tt}-g^{xx}K^{2}\right)  \omega^{2}%
+g^{yy}\partial_{y}^{2}-g^{zz}k_{z}^{2}+m^{2}\nonumber\\
&  +g^{xx}\left[  \left(  \frac{d}{dx}\ln f\right)  ^{2}+\frac{d^{2}}{dx^{2}%
}\ln f\right]  \nonumber\\
&  -2i\omega\left(  \frac{d}{dx}\ln f\right)  g^{xx}\left(  \frac{g^{tx}%
}{g^{xx}}-K\right)  .
\end{align}
We can eliminate the imaginary term, linear in $\omega$, by choosing%
\begin{equation}
K\equiv\frac{g^{tx}}{g^{xx}}~.\label{eq:K}%
\end{equation}

Let us assume a local Cartesian frame centered in one of the plates and two
different orientations of the Casimir apparatus. The first with the $x$ axis
being perpendicular to the plates, which we call the $x$-orientation, note
that this is the orientation considered in Ref. \cite{Sorge,zhang}. The second
one with the $y$ axis being perpendicular to the plates, which we call the
$y$-orientation. In what follows we call $\xi$ the coordinate perpendicular to
the plates, then we can generally refer to the $\xi$-orientation.

We fix a Dirichlet boundary condition in the first plate. Then we have%
\begin{align}
x\text{-orientation}:  &  f\equiv\sin(k_{x}x)~,\ g\equiv\exp\left(
ik_{y}y\right)  ~,\nonumber\\
&  k_{x}\equiv k_{\xi}~,\ k_{y}\in\mathbb{R}~,\nonumber\\
y\text{-orientation}:  &  f\equiv\exp\left(  ik_{x}x\right)  ~,\ g\equiv
\sin\left(  k_{y}y\right)  ~,\nonumber\\
&  k_{x}\in\mathbb{R~},\ k_{y}\equiv k_{\xi}~, \label{bc}%
\end{align}
where%
\begin{equation}
k_{\xi}\equiv\frac{\pi}{L}\left[  n-\frac{b}{2}\right]  ~,\ n=1,2,\ldots\ ,
\label{k-xi}%
\end{equation}
and$\ L$ is the (coordinate) distance between the plates. The parameter $b$
fixes the boundary condition on the second plate. For $b=0$, the same
condition as Dirichlet is fixed on the second plate, while for $b=1$, one has
a Neumann boundary condition on this second plate. The second case ($b=1$) is
known as a mixed (or hybrid) boundary condition.

Noting, however, that for both orientations $d^{2}g/dy^{2}=-k_{y}^{2}g$ and
that
\begin{equation}
\left(  \frac{d}{dx}\ln f\right)  ^{2}+\frac{d^{2}}{dx^{2}}\ln f=-k_{x}^{2}~,
\end{equation}
we can write
\begin{align}
\hat{D}  &  =\left[  \frac{\left(  g^{tx}\right)  ^{2}}{g^{xx}}-g^{tt}\right]
\omega^{2}-g^{xx}k_{x}^{2}\nonumber\\
&  -g^{zz}k_{z}^{2}-g^{yy}k_{y}^{2}+m^{2}~.
\end{align}
Therefore, for both orientations, the spectrum has the form%
\begin{equation}
\omega^{2}=\left(  g^{xx}k_{x}^{2}+g^{zz}k_{z}^{2}+g^{yy}k_{y}^{2}%
-m^{2}\right)  \left[  \frac{\left(  g^{tx}\right)  ^{2}}{g^{xx}}%
-g^{tt}\right]  ^{-1}~. \label{f2}%
\end{equation}
Note that, according to (\ref{bc}), $k_{x,y}$ can assume discrete or
continuous values.

The solutions must be normalized according to the KG scalar product
\begin{equation}
\left\langle \psi_{m},\psi_{n}\right\rangle =i\int_{\Sigma}\left(  \psi
_{n}^{\ast}\partial_{\mu}\psi_{m} -\psi_{m} \partial_{\mu}\psi_{n}^{\ast}
\right)  \,\sqrt{-g_{\Sigma}}\,n^{\mu}~d\Sigma~. \label{eq:SP}%
\end{equation}
Where $\Sigma$ is a spacelike Cauchy surface, $g_{\Sigma}$ is the determinant
of the metric induced in $\Sigma$ and $n^{\mu}$ is a timelike future-directed
unit vector orthogonal to $\Sigma$.

From the arc length, Eq.~(\ref{eq:LinEleCon}), we can calculate the inverse
metric
\begin{align}
\frac{\partial^{2}}{\partial s^{2}}  &  =\frac{g_{xx}}{\tilde{g}}%
\frac{\partial^{2}}{\partial t^{2}}-\frac{2g_{tx}}{\tilde{g}}\frac
{\partial^{2}}{\partial t\partial x}+\frac{g_{tt}}{\tilde{g}}\frac
{\partial^{2}}{\partial x^{2}}\nonumber\\
&  +\frac{1}{g_{yy}}\frac{\partial^{2}}{\partial y^{2}}+\frac{1}{g_{zz}}%
\frac{\partial^{2}}{\partial z^{2}}~. \label{invmetric}%
\end{align}
From where we can write (\ref{f2}) as
\begin{equation}
\omega^{2}=-g_{tt}\left(  \frac{g_{tt}}{\tilde{g}}k_{x}^{2}+\frac{1}{g_{yy}%
}k_{y}^{2}+\frac{1}{g_{zz}}k_{z}^{2}-m^{2}\right)  ~. \label{f}%
\end{equation}
The orthonormal vector to the $\Sigma$ surface can be constructed as
\begin{align}
\frac{\partial^{2}}{\partial s^{2}}  &  =\frac{1}{\tilde{g}}\left[  \left(
\sqrt{g_{xx}}\frac{\partial}{\partial t}-\frac{g_{tx}}{\sqrt{g_{xx}}}%
\frac{\partial}{\partial x}\right)  ^{2}-\left(  \frac{(g_{tx})^{2}}{g_{xx}%
}-g_{tt}\right)  \frac{\partial^{2}}{\partial x^{2}}\right] \nonumber\\
&  +\frac{1}{g_{yy}}\frac{\partial^{2}}{\partial y^{2}}+\frac{1}{g_{zz}}%
\frac{\partial^{2}}{\partial z^{2}}~.
\end{align}
From this expression, we find
\begin{equation}
n^{\mu}=\left(  \sqrt{\frac{g_{xx}}{\tilde{g}}},\frac{g_{tx}}{\sqrt{\tilde
{g}g_{xx}}},0,0\right)  ~,\ g_{\Sigma}=\frac{g}{g_{tt}}~. \label{eq:NorVec}%
\end{equation}
As stated, this is a timelike normalized vector, $n^{\mu}n_{\mu}=1$. Using the
above $n^{\mu}$ vector in (\ref{eq:SP}), we find the normalization of the
solutions of the KG equation%
\begin{equation}
\left\vert N\right\vert ^{2}=-\frac{g_{tt}\sqrt{-g_{tt}g_{xx}g_{yy}g_{zz}}%
}{g\left(  2\pi\right)  ^{2}L}\left[  \omega+\frac{g_{tt}g_{tx}}{\tilde{g}%
}G_{\xi}k_{x}\right]  ^{-1}~, \label{eq:Nn1}%
\end{equation}
where%
\begin{equation}
G_{\xi}=%
\begin{cases}
0~, & \text{for }x\text{-orientation}\\
1~, & \text{for }y\text{-orientation }%
\end{cases}
~.
\end{equation}

The main consequence of our development comes from the dependence of the
normalization factor, Eq. (\ref{eq:Nn1}), with the orientation of the plates.
In the $x$-orientation, the discrete modes of the field, which do not
contribute to the normalization in Eq. (\ref{eq:SP}), are those influenced by
the non-diagonal part of the metric. Hence, the non-diagonal metric does not
affect the normalization and we have the usual $1/\omega$ dependence, present
in the case of the diagonal metric (as well as in the case of the flat space).
However, in the $y$-orientation, the continuous modes of the field, which do
contribute to the normalization, are influenced by the non-diagonal term in
the metric. Therefore $N$ becomes dependent on $g_{tx}$, which, as we will
see, also affect the vacuum energy and consequently the Casimir energy.

Regarding this dependence of the normalization with the orientation, it is
interesting to note that, in the case of Casimir apparatus in the format of a
box (see, e.g. \cite{Mostepanenko}), in which case all modes of the field are
discrete, the normalization of the field is insensitive to the orientation,
because these modes do not contribute to the KG product, regardless of the
non-diagonal term in the metric.

\section{\label{casimir}The vacuum and Casimir energy}

In order to compute the Casimir energy, we proceed with the usual techniques
of quantum fields in curved spacetimes \cite{Birrell}. From the
energy-momentum tensor for the field,%
\begin{equation}
T_{\mu\nu}=\partial_{\mu}\psi\partial_{\nu}\psi^{\ast}-\frac{1}{2}g_{\mu\nu
}\left(  g^{\rho\sigma}\partial_{\rho}\psi\partial_{\sigma}\psi^{\ast}%
-m^{2}\left\vert \psi\right\vert ^{2}\right)  ~, \label{eq:Tmunu}%
\end{equation}
we evaluate the average energy density of the vacuum $\bar{\epsilon}_{vac}$
for the scalar field within the cavity (Casimir's energy density). This
average value reads
\begin{equation}
\bar{\epsilon}_{vac}=\frac{1}{V_{p}}\int dxdydz\sqrt{-g_{\Sigma}}%
\,\epsilon_{vac}~, \label{a}%
\end{equation}
where%
\begin{equation}
V_{P}=\int dxdydz\sqrt{-g_{\Sigma}}~, \label{vp}%
\end{equation}
is the proper volume of the cavity, as measured by static observer with four
velocity
\begin{equation}
w^{\mu}=\frac{1}{\sqrt{g_{tt}}}\delta^{\mu_{t}}~,\ g_{tt}>0~,
\label{eq:VelStaObs}%
\end{equation}
while
\begin{align}
\epsilon_{vac}  &  =w^{\mu}w^{\nu}\left\langle 0\right\vert T_{\mu\nu
}\left\vert 0\right\rangle \nonumber\\
&  =\left(  g_{tt}\right)  ^{-1}\sum_{n}\int dk_{\alpha}dk_{z}T_{tt}\left[
\psi(\vec{k}),\psi^{\ast}(\vec{k})\right]  ~. \label{eq:VacEne-2}%
\end{align}
The bilinear form $T_{\mu\nu}$ was defined in (\ref{eq:Tmunu}). From now on,
we use $\alpha$ to denote the direction orthogonal to the orientation $\xi
$\ of the plates\ and $z$ direction, i.e., $\alpha=x$ for the $y$-orientation
and $\alpha=y$ for the $x$-orientation.

Our goal is to find the gravity-induced corrections to the vacuum energy
density for a scalar quantum field enclosed in the cavity. From (\ref{a}) and
(\ref{eq:VacEne-2}) we have,%
\begin{equation}
\bar{\epsilon}_{vac}=\frac{1}{g_{tt}}\sum_{n}\int dk_{\alpha}dk_{z}\frac
{1}{V_{p}}\int d\Sigma\sqrt{-g_{\Sigma}}T_{tt}\left[  \psi(\vec{k}),\psi
^{\ast}(\vec{k})\right]  ~. \label{c}%
\end{equation}
Using (\ref{invmetric})\ to determine $g^{\rho\sigma}\partial_{\rho}%
\psi\partial_{\sigma}\psi^{\ast}$, and the solutions (\ref{eq:SolKG}), with
the appropriate choice (\ref{bc}), we find%
\begin{align}
&  \int_{\Sigma}T_{tt}\sqrt{-g_{\Sigma}}d\Sigma=\int_{z}\int_{\alpha}\int
_{\xi}T_{tt}\left(  \psi,\psi^{\ast}\right)  \sqrt{-g_{\Sigma}}~dzd\alpha
d\xi~,\nonumber\\
&  T_{tt}\left(  \psi,\psi^{\ast}\right)  =-g_{tt}\left\vert N\right\vert
^{2}\nonumber\\
&  \times\left[  \sin^{2}\left(  k_{\xi}\xi\right)  \left(  g^{zz}k_{z}%
^{2}+g^{\alpha\alpha}k_{\alpha}^{2}-m^{2}\right)  +\frac{g^{\xi\xi}k_{\xi}%
^{2}}{2}\right]  ~,
\end{align}
where $k_{\alpha}\in%
\mathbb{R}
$ and $\xi$ the variable along the orientation of the plates,
\begin{align}
x\text{-orientation}  &  :\xi=x\in\left[  0,L\right]  ~,\ \alpha
=y~,\nonumber\\
y\text{-orientation}  &  :\xi=y\in\left[  0,L\right]  ~,\ \alpha=x~.
\end{align}
Using (\ref{f}), and evaluating the integrals, we find%
\begin{equation}
\frac{1}{V_{p}}\int_{\Sigma}T_{tt}(\psi,\psi^{\ast})\sqrt{-g_{\Sigma}}%
d\Sigma=\frac{\left\vert N\right\vert ^{2}\omega^{2}}{2}~. \label{b}%
\end{equation}
Note that, apart from $N$, the result above does not depend on the orientation
of the apparatus. Using (\ref{b}) in (\ref{c}) we have%
\begin{equation}
\bar{\epsilon}_{vac}=(g_{tt})^{-1}\sum_{n}\int dk_{\alpha}dk_{z}%
\frac{\left\vert N\right\vert ^{2}\omega^{2}}{2}~. \label{eq:MeaVacEne}%
\end{equation}

Substituting now Eq. (\ref{eq:Nn1}) for the normalization, we obtain,
\begin{equation}
\left\vert N\right\vert ^{2}\omega^{2}=-\frac{\omega^{2}g_{tt}\sqrt
{-g_{tt}g_{xx}g_{yy}g_{zz}}}{g\left(  2\pi\right)  ^{2}L\left(  \omega
+\frac{g_{tt}g_{tx}}{\tilde{g}}G_{\xi}k_{x}\right)  }~.
\end{equation}
Note that, once $G_{\xi}=0$ for the $x$-orientation, we can write $k_{x}$ or
$k_{\alpha}$ in the above equation. We can now determine the average vacuum
energy density (\ref{c}),
\begin{equation}
\bar{\epsilon}_{vac}=-\frac{\sqrt{-g_{tt}g_{xx}g_{yy}g_{zz}}}{2g\left(
2\pi\right)  ^{2}L}\sum_{n}I_{n}~, \label{evm}%
\end{equation}
where, making explicit the discrete dependence of $N$ and $\omega$ due to
(\ref{k-xi}), $k_{\xi}=\left(  k_{\xi}\right)  _{n}$,
\begin{equation}
I_{n}=\int dk_{\alpha}dk_{z}\frac{\omega_{n}^{2}}{\omega_{n}+\frac
{g_{tt}g_{tx}}{\tilde{g}}G_{\xi}k_{\alpha}}~, \label{eq:I-1}%
\end{equation}
and $\omega_{n}$ is given in (\ref{f}), or (\ref{f2}). Making the variables
transformations
\begin{equation}
k_{\alpha}=\sqrt{-\frac{1}{g^{\alpha\alpha}}}k\cos\theta,\quad k_{z}%
=\sqrt{-g_{zz}}k\sin\theta~,
\end{equation}
we find
\begin{equation}
\omega_{n}^{2}=\left[  k^{2}-g^{\xi\xi}k_{\xi}^{2}+m^{2}\right]  g_{tt}~.
\label{eq:Omen}%
\end{equation}
Substituting (\ref{eq:Omen}) in (\ref{eq:I-1}), and making the integral in
$\theta$, we have%
\begin{equation}
I_{n}=\frac{\pi}{F^{2}}\sqrt{\frac{g_{tt}g_{zz}}{g^{\alpha\alpha}}}\left[
I_{+}+\left(  1-F\right)  \left(  g^{\xi\xi}k_{\xi}^{2}-m^{2}\right)
I_{-}\right]  ~, \label{in}%
\end{equation}
where%
\begin{equation}
u=Fk^{2}-g^{\xi\xi}k_{\xi}^{2}+m^{2}~,\ F=1+\frac{g_{tt}}{g^{\alpha\alpha}%
}\left(  \frac{g_{tx}}{\tilde{g}}G_{\xi}\right)  ^{2}~, \label{eq:u}%
\end{equation}
and%
\begin{equation}
I_{\pm}=\int_{u(0)}^{u(\infty)}u^{\pm1/2}~du~. \label{in2}%
\end{equation}

Both integrals $I_{\pm}$ in (\ref{in}) diverge, but can be regularized using,
for example, zeta regularization \cite{zeta}. Following the usual method, we
start by considering the integral,%
\begin{equation}
\int_{u\left(  0\right)  }^{\infty}u^{-s/2}~du=-\frac{u\left(  0\right)
^{1-s/2}}{1-s/2},\quad\operatorname{Re}\left(  s\right)  >2~. \label{i1}%
\end{equation}
Where, from (\ref{eq:u}), we have
\begin{equation}
u\left(  0\right)  =u_{n}\left(  0\right)  =-\left(  g^{\xi\xi}k_{\xi}%
^{2}-m^{2}\right)  ~,
\end{equation}
with $k_{\xi}$ giving in (\ref{k-xi}). Relaxing for a moment the restriction
$\operatorname{Re}\left(  s\right)  >2$ in (\ref{i1}), we can write (\ref{in})
as%
\begin{align}
I_{n}  &  =\left[  \left(  1-F\right)  u_{n}\left(  0\right)  \left.
\frac{u_{n}\left(  0\right)  ^{1-s/2}}{1-s/2}\right\vert _{s=1}-\left.
\frac{u_{n}\left(  0\right)  ^{1-s/2}}{1-s/2}\right\vert _{s=-1}\right]
\nonumber\\
&  \times\frac{\pi}{F^{2}}\sqrt{\frac{g_{tt}g_{zz}}{g^{\alpha\alpha}}}~,
\end{align}
or yet,%
\begin{equation}
I_{n}=\frac{2\pi}{F^{2}}\sqrt{\frac{g_{tt}g_{zz}}{g^{\alpha\alpha}}}\left(
\frac{2}{3}-F\right)  \left[  u_{n}\left(  0\right)  \right]  ^{3/2}~.
\end{equation}
Now we can evaluate the (analytic continuation of the)\ sum%
\begin{align}
&  \sum_{n}u_{n}\left(  0\right)  ^{3/2}=\pi^{3}\left[  \frac{\sqrt{-g^{\xi
\xi}}}{L}\right]  ^{3}\sum_{n}\left[  \left(  n+\frac{b}{2}\right)  ^{2}%
+q^{2}\right]  ^{3/2}~,\nonumber\\
&  q^{2}=-\frac{L^{2}m^{2}}{\pi^{2}g^{\xi\xi}}~,
\end{align}
and use the relation \cite{zeta}%
\begin{align}
&  \sum_{n=-\infty}^{\infty}\left[  \left(  n+\frac{b}{2}\right)  ^{2}%
+q^{2}\right]  ^{-s}=\sqrt{\pi}\frac{\Gamma\left(  s-\frac{1}{2}\right)
}{\Gamma\left(  s\right)  }\nonumber\\
&  +\frac{4\pi^{s}}{\Gamma\left(  s\right)  }q^{\frac{1}{2}-s}\sum
_{n=1}^{\infty}\left(  -1\right)  ^{bn}n^{s-\frac{1}{2}}K_{s-\frac{1}{2}%
}\left(  2\pi nq\right)  ~, \label{4.13}%
\end{align}
where $K_{\nu}$ is the modified Bessel function of the second kind. As a
result
\begin{align}
\sum_{n=0}^{\infty}u_{n}\left(  0\right)  ^{3/2}  &  =\frac{3q^{2}\pi}%
{2}\left[  \frac{\sqrt{-g^{\xi\xi}}}{L}\right]  ^{3}\sum_{n=1}^{\infty}%
\frac{\left(  -1\right)  ^{bn}}{n^{2}}K_{2}\left(  2\pi nq\right) \nonumber\\
&  +E_{0}\,, \label{s0}%
\end{align}
where we use $K_{-2}\left(  x\right)  =K_{2}\left(  x\right)  $ and%
\begin{equation}
E_{0}=\frac{3}{8}\left[  \frac{\pi\sqrt{-g^{\xi\xi}}}{L}\right]  ^{3}%
\Gamma\left(  -2\right)  ~.
\end{equation}
From the asymptotic behavior of the Bessel function \cite{AbraSt1072}%
\begin{equation}
x>>n\Rightarrow K_{n}\left(  x\right)  \sim\sqrt{\pi}\frac{e^{-x}}{\sqrt{2x}%
}~,
\end{equation}
we see that the divergent term $E_{0}$ can be associated with the limit
$L\rightarrow\infty$ ($q\rightarrow\infty$)\ and, consequently, it corresponds
to the (always divergent) vacuum energy without boundaries. This term must be
discounted in the computation of the Casimir energy \cite{EliRo1989}.
Therefore, discounting $E_{0}$,
\begin{align}
\sum_{n=0}^{\infty}I_{n}  &  =\frac{3q^{2}\pi^{2}}{F^{2}}\sqrt{\frac
{g_{tt}g_{zz}}{g^{\alpha\alpha}}}\left[  \frac{\sqrt{-g^{\xi\xi}}}{L}\right]
^{3}\left(  \frac{2}{3}-F\right) \nonumber\\
&  \times\sum_{n=1}^{\infty}\frac{\left(  -1\right)  ^{bn}}{n^{2}}K_{2}\left(
2q\pi n\right)  ~. \label{l}%
\end{align}
For our consideration of zero order expansion of the metric in the region of
the apparatus, the term in brackets from the above equation can be recognized
as the proper length $L_{p}$ \cite{Sorge},%
\begin{equation}
L_{p}=\int_{0}^{L}\sqrt{-\frac{1}{g^{\xi\xi}}}~d\xi=L\sqrt{-\frac{1}{g^{\xi
\xi}}}~. \label{lp}%
\end{equation}
Finally, after some manipulations of the elements of the metric, using
(\ref{lp}), (\ref{l}) and (\ref{eq:u}) in (\ref{evm}), we can write the
Casimir energy in the $\xi$-direction as%
\begin{equation}
\bar{\epsilon}_{vac}^{\left(  \xi\right)  }=\left(  \frac{\tilde{g}}%
{g_{tt}g_{xx}}\right)  ^{\left(  4G_{\xi}-1\right)  /2}\left(  1+3G_{\xi}%
\frac{g_{tx}^{2}}{\tilde{g}}\right)  \mathcal{E}_{m}\mathcal{~},
\label{final-m}%
\end{equation}
where,%
\begin{equation}
\mathcal{E}_{m}=-\frac{m^{2}}{8\pi^{2}L_{p}^{2}}\sum_{n=1}^{\infty}%
\frac{\left(  -1\right)  ^{bn}}{n^{2}}K_{2}\left(  2mL_{p}n\right)  ~.
\label{Em}%
\end{equation}
For the Dirichlet boundary condition ($b=0$) $\mathcal{E}_{m}$ is the
well-known Casimir energy for a massive scalar field in flat spacetime,
calculated using cutoff or dimensional regularization methods
\cite{PluMu1986,AmbWo1983}. Although the result is not surprising, we have not
been able to find this result in the literature for a massive scalar field
with mixed boundary conditions ($b=1$).

Equation (\ref{final-m}) is our main result. This expression is unchanged by
$g_{yy}\leftrightarrow g_{zz}$, indicating that a reorientation of the plates
in the $z$-direction reproduces the same result (with $G_{\xi}\equiv G_{y}%
=1$). Then, while the $x$-orientation has no rotation symmetry, we expect the
$y$-orientation to have a symmetry for rotations on the $x$ axis. As
highlighted in \cite{Sorge} the new effects, not present on the flat
background, are due to the breaking of $x\rightarrow-x$ symmetry (azimuthal
reflection in Kerr geometry). Although diagonalization of Eq.
(\ref{eq:LinEleCon}) is possible, it would hide this symmetry breaking thus
eliminating the new effects originally associated with space-time described by
Eq. (\ref{st-axi-sym}). However, it is possible that the transformation of the
spatial coordinates from (\ref{st-axi-sym}) to (\ref{eq:LinEleCon}) is
precisely the one that restores this symmetry, diagonalizing the metric of the
local Cartesian coordinate frame (e.g., the zero angular momentum observer in
Kerr geometry). In this special case, we have%
\begin{equation}
g_{tx}=0\Rightarrow\frac{g_{tt}g_{xx}}{\tilde{g}}=1\Rightarrow\bar{\epsilon
}_{vac}^{\left(  \xi\right)  }=\mathcal{E}_{m}\mathcal{~},
\end{equation}
and all corrections induced by gravity disappear. We will present some
concrete examples.

Finally, we highlight the last term in parenthesis in (\ref{final-m}). This
term could cause a change in the Casimir energy sign, without the insertion of
a medium or a change in the boundary condition. A new and unexpected effect.
Namely, this change occurs if%
\begin{equation}
g_{tx}^{2}<-G_{\xi}\frac{g_{tt}g_{xx}}{2}~. \label{cond}%
\end{equation}
To see that this condition can actually be met, in the next sections we
analyze some specific geometries.

This possible change of sign in the Casimir energy can be understood in
connection to the dependence of normalization, Eq. (\ref{eq:Nn1}), on $g_{tx}$
as follows. In the $y$-orientation, the normalization of the field, and
consequently the vacuum energy, is affected by the non-diagonal term in the
metric. In this case, the balance of energy between continuous and discrete
modes inside the apparatus can be modified by the spacetime, changing the sign of the
Camisir energy. Conversely, in the $x$-direction, the normalization is not
affected by the non-diagonal term, hence the balance of energy inside the
apparatus can not be changed.

Now, let's consider the case without mass. The expressions for the massless
scalar field can be obtained from the limit $m\rightarrow0$. For this goal, we
use the behavior of the Bessel functions for small arguments \cite{AbraSt1072}%
\begin{equation}
z\rightarrow0\Rightarrow K_{\nu}\left(  z\right)  \sim\frac{1}{2}\Gamma\left(
\nu\right)  \left(  \frac{z}{2}\right)  ^{-\nu}.
\end{equation}
Which implies%
\begin{equation}
\mathcal{E}_{m\rightarrow0}\sim\frac{-1}{16\pi^{2}L_{p}^{4}}\sum_{n=1}%
^{\infty}\frac{\left(  -1\right)  ^{bn}}{n^{4}}=-\left(  -\frac{7}{8}\right)
^{b}\frac{\pi^{2}}{1440L_{p}^{4}}~. \label{E}%
\end{equation}
We can recognize $\mathcal{E}\equiv\mathcal{E}_{m\rightarrow0}$\ as the
Casimir energy in the flat spacetime. For $b=1$ we have the $(-7/8)$ factor,
resulting a repulsive effect. For the massless case, this sign change,
resulting from a mixed boundary condition, is a well-known effect
\cite{Mostepanenko}.

The expression (\ref{final-m})\ reproduces the results in \cite{zhang} by
choosing the $x$-orientation ($G_{\xi}\equiv G_{x}=0$) and making
$m\rightarrow0$. Unlike flat spacetime, when condition (\ref{cond}) is
satisfied, we have a repulsive Casimir force for the same boundary condition
in both plates ($b=0$), and an attractive force for a mixed boundary condition
($b=1$).

\section{\label{cylinder}Constant linear momentum cylinder}

As a first example, we consider the spacetime external to a distribution of
mass-energy with cylindrical symmetry. The distribution is in a non-rotating
stationary state of motion along the symmetry axis $\tilde{x}$. Such a system
can be described by the metric \cite{GeliTi2000}
\begin{align}
ds^{2}  &  =r^{2q_{-}}\cos\left(  2k\ln r\right)  \left(  dt^{2}-d\tilde
{x}^{2}\right)  -dr^{2}\nonumber\\
&  -2r^{2q_{-}}\sin\left(  2k\ln r\right)  dtd\tilde{x}-r^{2q_{+}}d\theta
^{2}~, \label{em}%
\end{align}
where $k$ is a (not necessary positive) constant related with the source
momentum, and%
\begin{equation}
q_{\pm}=\frac{1}{3}\left[  1\pm2\left(  1+3k^{2}\right)  ^{\frac{1}{2}%
}\right]  ~.
\end{equation}
This metric is stationary but can be static when $k=0$. For $k\neq0$ it's an
example of a metric satisfying the conditions which define Eq.
(\ref{st-axi-sym}). Possible physical sources for this metric are discussed in
\cite{GeliTi2000}.

\begin{figure}
[ptb]
\begin{center}
\includegraphics[
height=2.1482in,
width=2.5884in
]%
{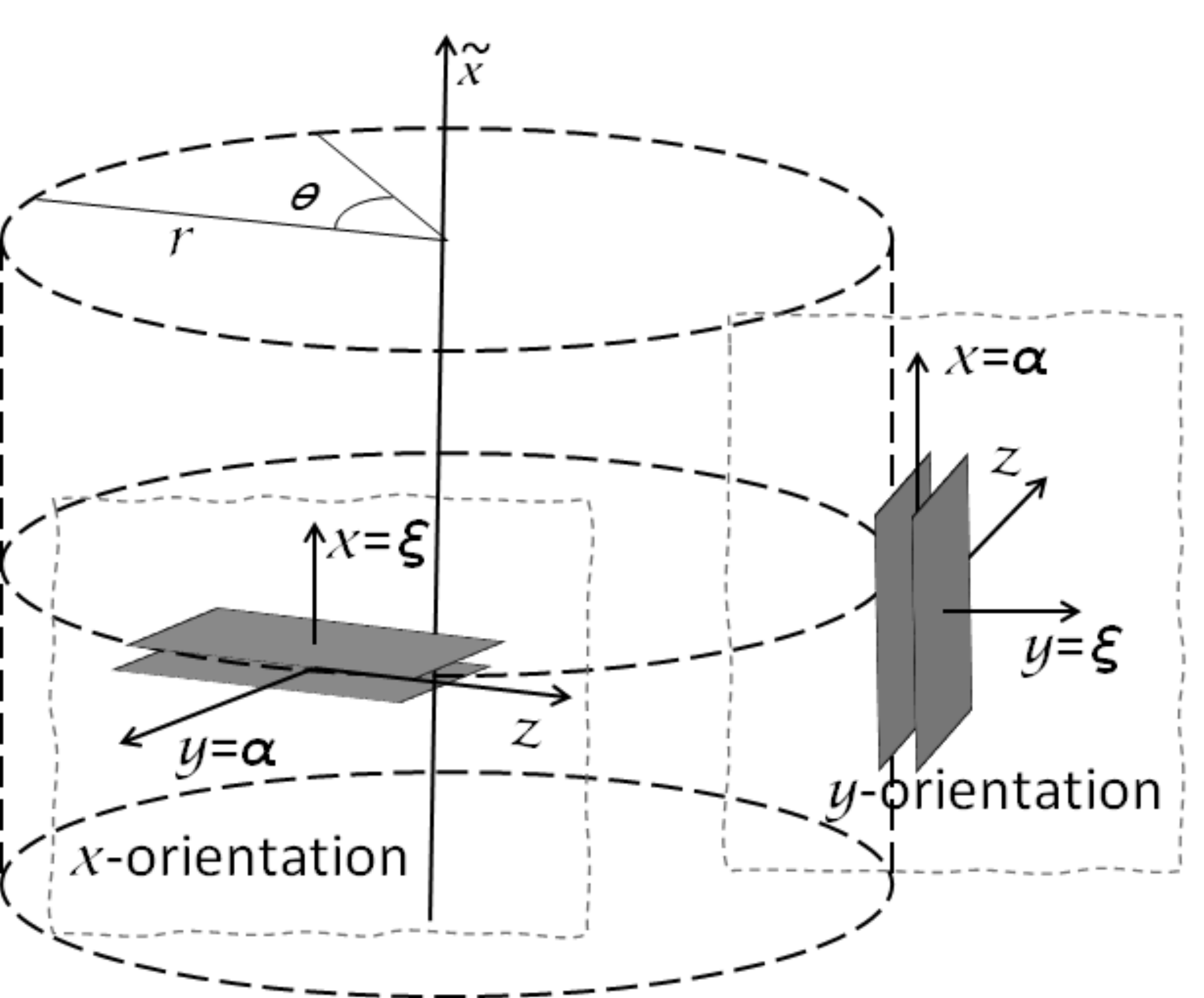}%
\caption{Orientations of the Casimir apparatus for the cylindrically symmetric
distribution.}%
\label{fig2}%
\end{center}
\end{figure}

We want to consider a Casimir apparatus moving along the $\tilde{x}$
direction, with constant $r$, $\theta$ and velocity $v=d\tilde{x}/dt$. This
apparatus has four-velocity%
\begin{equation}
\tilde{w}^{\mu}=S\left(  k,r,v\right)  \left(  1,v,0,0\right)  ~,
\end{equation}
where
\begin{equation}
S^{-1}\left(  k,r,v\right)  =r^{q_{-}}\sqrt{\left(  v^{2}-1\right)
\cos\left(  2k\ln r\right)  +2v\sin\left(  2k\ln r\right)  }~.
\end{equation}
The considered orientations are illustrated in Fig. \ref{fig2}.%

In order to change for a comoving Cartesian reference frame, we first consider
the transformation,%
\begin{equation}
x^{\prime}=\tilde{x}-vt~.
\end{equation}
Next, we consider a Cartesian local frame $\left(  x,y,z\right)  $ attached to
the Casimir device and centered on one of the plates. In this frame%
\begin{equation}
dx=dx^{\prime},~dy=dr~,~dz=rd\theta~.
\end{equation}
In the comoving frame, the metric assumes the form,%
\begin{align}
&  g_{tt}=2vg_{tx}-\left(  1+v^{2}\right)  g_{xx}~,\ g_{yy}=-1~,\nonumber\\
&  g_{tx}=g_{xx}v-r^{2q_{-}}\sin\left(  2k\ln r\right)  ~,\nonumber\\
&  g_{xx}=-r^{2q_{-}}\cos\left(  2k\ln r\right)  ~,\ g_{zz}=-r^{2\left(
q_{+}-1\right)  }~.
\end{align}
In this new metric the apparatus is static with four velocity%
\begin{equation}
w^{\mu}=S(k,r,v)(1,0,0,0)~.
\end{equation}
In addition, for
\begin{equation}
v=v_{d}\equiv-\tan\left(  2k\ln r\right)  ~,
\end{equation}
the three-velocity $\mathbf{w}=0$ is static in coordinates in which the metric
locally takes a diagonal form $g_{tx}=0$. It corresponds to an observer (or
apparatus) with zero linear momentum. This zero linear momentum observer, with
non-vanishing velocity with respect to the original metric (\ref{em}),
represents a form of \textquotedblleft frame dragging\textquotedblright, as
the one associated with the spacetime of sources endowed with rotation. In
other words, using the elements of the original metric (\ref{em}),
\begin{equation}
v_{d}=-\frac{g_{t\tilde{x}}}{g_{\tilde{x}\tilde{x}}}~,
\end{equation}
is the dragging linear velocity of spacetime.

As pointed out in \cite{GeliTi2000,ThaMo1998}, for a fixed $\theta$, the
Killing vectors $\partial_{t}$ and $\partial_{x}$ may interchange their
spacelike/timelike characteristic. Nevertheless, it is possible to define a
time orientation at each spacetime point (except $r=0$). Since we consider a
fixed $r$, this time orientation does not change. Then, to preserve the time
orientation, we must set%
\begin{equation}
v_{-}<v<v_{+}~,\ v_{\pm}=v_{d}\pm\sqrt{v_{d}^{2}+1}~. \label{range}%
\end{equation}
The $v_{\pm}$ values represent the limit velocities of the apparatus, which we
refer to as the ultra-relativistic cases.

Substituting the above components of the metric in (\ref{final-m}), and
choosing the $x$-orientation for the apparatus ($G_{\xi}=0$), we have
\begin{equation}
\bar{\epsilon}_{vac}^{\left(  x\right)  }=\mathcal{E}_{m}\sqrt{1-\frac{\left(
v-v_{d}\right)  ^{2}}{v_{d}^{2}+1}}~.
\end{equation}
While for the $y$-orientation we have%
\begin{equation}
\bar{\epsilon}_{vac}^{\left(  y\right)  }=\mathcal{E}_{m}\frac{v_{d}%
^{2}+1-3\left(  v-v_{d}\right)  ^{2}}{\left[  \left(  v_{d}^{2}+1\right)
-\left(  v-v_{d}\right)  ^{2}\right]  ^{\frac{3}{2}}}\sqrt{v_{d}^{2}+1}~.
\end{equation}
The values coincide for the zero linear momentum%
\begin{equation}
v=v_{d}\Rightarrow\bar{\epsilon}_{vac}^{\left(  y\right)  }=\bar{\epsilon
}_{vac}^{\left(  x\right)  }=\mathcal{E}_{m}~.
\end{equation}
But it behaves completely differently for all other velocities. In particular,
in the ultra-relativistic regimes, $v\rightarrow v_{\pm}$ , we have,
\begin{equation}
v\rightarrow v_{\pm}\Rightarrow\left\{
\begin{array}
[c]{l}%
\bar{\epsilon}_{vac}^{\left(  x\right)  }\rightarrow0\\
\bar{\epsilon}_{vac}^{\left(  y\right)  }\rightarrow\left(  \pm\right)  \infty
\end{array}
\right.  ~,
\end{equation}
with the $\left(  \pm\right)  $ sign for Dirichlet and mixed boundary
condition, respectively. While in the $x$-orientation the Casimir energy goes
to zero, in the $y$-orientation this energy diverges. Remembering that, as in
the Minkowski case, the $\mathcal{E}_{m}$ energy decays with the increase in
the mass, in the $y$-orientation a very massive scalar field could still
produce a Casimir force. Besides, in the $y$-orientation, the energy, not only
diverges, but with a sign opposite to $\mathcal{E}_{m}$. So, without changing the
boundary condition, we can change the Casimir force from attractive to
repulsive. Namely, the Casimir energy assumes the usual intensity, but with
opposite sign when%
\begin{align}
&  v=\hat{v}_{\pm}\Rightarrow\bar{\epsilon}_{vac}^{\left(  y\right)
}=-\mathcal{E}_{m}~~,\nonumber\\
&  \hat{v}_{\pm}=v_{d}\pm\sqrt{\left(  2\sqrt{3}-3\right)  \left(  v_{d}%
^{2}+1\right)  }~.
\end{align}

The energy disappear at the velocity%
\begin{equation}
v=v_{0\pm}=v_{d}\pm\sqrt{\frac{v_{d}^{2}+1}{3}}\Rightarrow\bar{\epsilon}%
_{vac}^{\left(  y\right)  }=0~.
\end{equation}
Being attractive for $v_{0-}<v<v_{0+}$ and repulsive out this interval.

It is important to note that all values of $\hat{v}_{\pm}$ and $v_{0\pm}$ are
in the range (\ref{range}). This means that the effect of the change in the
Casimir force sign cannot be associated with any causal defect in the
trajectory, or other prohibited relativistic process.

\section{\label{kerr}Application for Kerr metric}

\begin{figure}
[ptb]
\begin{center}
\includegraphics[
height=1.644in,
width=3.474in
]%
{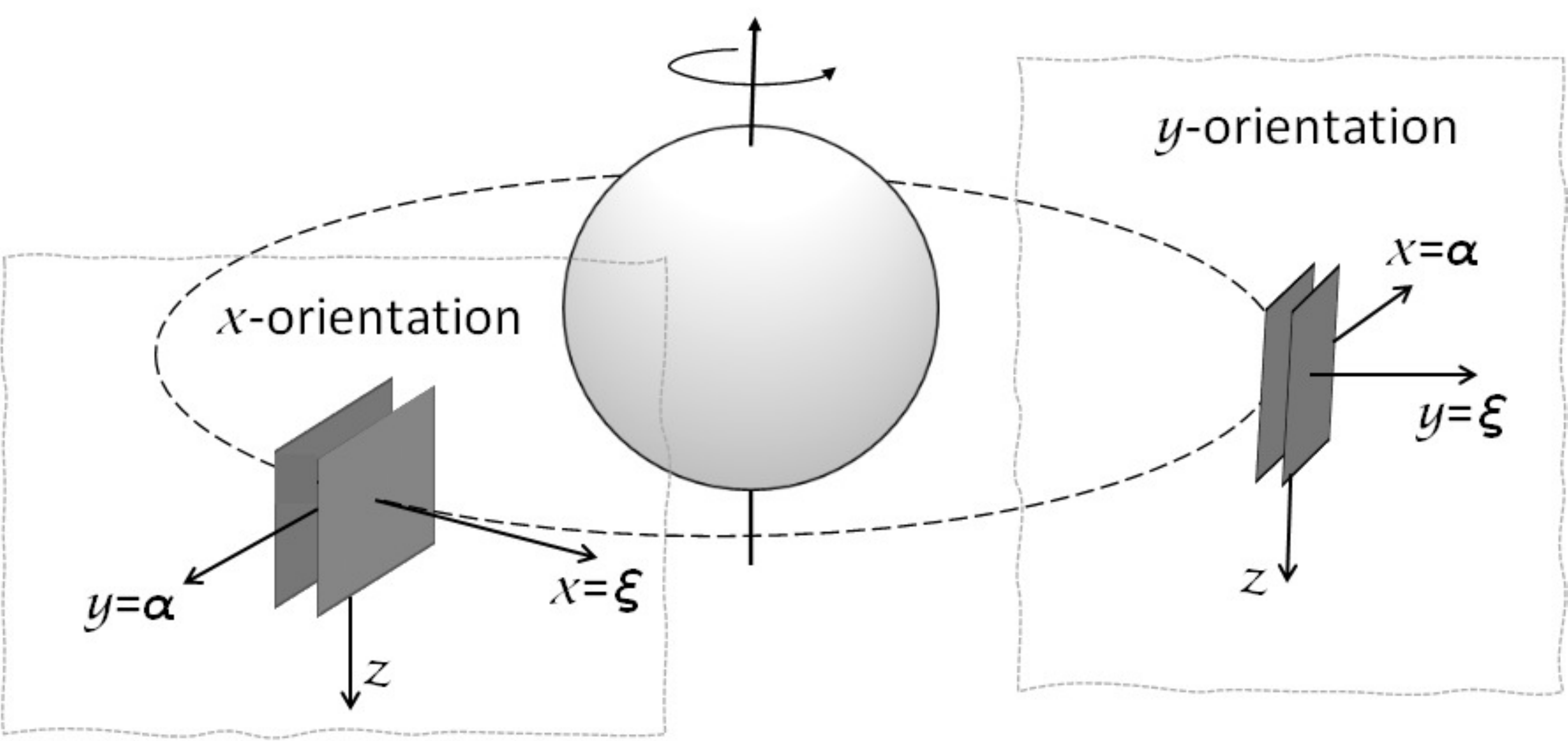}%
\caption{Orientations of the Casimir apparatus for the Kerr metric.}%
\label{orient}%
\end{center}
\end{figure}

As a second example, we now apply our result to Kerr's geometry. A case with
more direct applications in physical problems. In this case, the considered
orientations are illustrated in Fig. \ref{orient}.%

Following \cite{Sorge} we start with the Kerr metric in the Boyer-Lindquist
coordinates,%
\begin{align}
ds^{2}  &  =\left(  1-\frac{2Mr}{\Sigma}\right)  dt^{2}+2\frac{A}{\Sigma
}\omega_{d}\sin^{2}\theta dtd\varphi\nonumber\\
&  -\frac{\Sigma}{\Delta}dr^{2}-\Sigma d\theta^{2}-\frac{A\sin^{2}\theta
}{\Sigma}d\varphi^{2}~, \label{Kerr}%
\end{align}
where%
\begin{align}
&  \Sigma=r^{2}+a^{2}\cos^{2}\theta,\hspace{0.5cm}\Delta=r^{2}+a^{2}%
-2Mr~,\nonumber\\
&  A=\left(  r^{2}+a^{2}\right)  \Sigma+2Mra^{2}\sin^{2}\theta~,
\label{parameters}%
\end{align}
$a=J/M$ is the Komar angular momentum by unit mass, and%
\begin{equation}
\omega_{d}=-\frac{g_{t\varphi}}{g_{\varphi\varphi}}=\frac{2Mar}{A}~,
\end{equation}
is the dragging angular velocity of spacetime. We are interested in circular
equatorial orbits, so, as in the previous case, we consider a comoving
observer with the apparatus\ via the transformation,%
\begin{equation}
\varphi^{\prime}=\varphi-\Omega t~, \label{rf}%
\end{equation}
where $\Omega$ is the angular velocity of the Casimir apparatus. With the
transformation (\ref{rf}) the Kerr metric (\ref{Kerr}) becomes
\begin{align}
ds^{2}  &  =g_{tt}dt^{2}-\frac{\Sigma}{\Delta}dr^{2}-\Sigma d\theta^{2}%
-\frac{A\sin^{2}\theta}{\Sigma}d\varphi^{^{\prime}2}\nonumber\\
&  -2\left(  \Omega-\omega_{d}\right)  \frac{A}{\Sigma}\sin^{2}\theta
\,dtd\varphi^{\prime}~, \label{9}%
\end{align}
where%
\begin{equation}
g_{tt}=1-\frac{2A}{\Sigma}\left[  \frac{Mr}{A}-\Omega\left(  \omega_{d}%
-\frac{\Omega}{2}\right)  \sin^{2}\theta\right]  ~.
\end{equation}
For an apparatus in the equatorial orbit ($\theta=\pi/2$, not necessarily
geodesic) we can write
\begin{equation}
g_{tt}=\frac{\Delta\Sigma}{A}\left[  1-\frac{A^{2}}{\Delta\Sigma^{2}}\left(
\Omega-\omega_{d}\right)  ^{2}\right]  ~.
\end{equation}
Allowed observers require $g_{tt}>0$, so%
\begin{equation}
\Omega_{-}<\Omega<\Omega_{+}~,\ \Omega_{\pm}=\omega_{d}\pm\frac{\Sigma
\sqrt{\Delta}}{A}~. \label{pv}%
\end{equation}

Now we consider the comoving Cartesian local frame $\left(  x,y,z\right)  $
attached to the Casimir device and centered on one of the plates,
\begin{equation}
dx=rd\varphi^{\prime},~dy=dr,~dz=rd\theta~. \label{eq13}%
\end{equation}
As a result, the metric (\ref{9}) takes the form%
\begin{align}
g_{tx}  &  =-\frac{A}{r^{3}}\left(  \Omega-\omega_{d}\right)  ~,\ g_{xx}%
=-\frac{A}{r^{4}}~,\nonumber\\
g_{yy}  &  =-\frac{r^{2}}{\Delta}~,\ g_{zz}=-1~. \label{3.16}%
\end{align}

Substituting the above components of the metric in (\ref{final-m}), and
choosing the $x$-orientation ($G_{x}=0$),
\begin{equation}
\bar{\epsilon}_{vac}^{\left(  x\right)  }=R\left(  r;\Omega,M,a\right)
\mathcal{E}_{m}~,
\end{equation}
where%
\begin{align}
R\left(  r;\Omega,M,a\right)   &  =\sqrt{\frac{g_{tt}g_{xx}}{\tilde{g}}%
}\nonumber\\
&  =\left[  1-\frac{A^{2}}{r^{4}\Delta}\left(  \Omega-\omega_{d}\right)
^{2}\right]  ^{\frac{1}{2}}~.
\end{align}
That is the result obtained in \cite{Sorge} for the massless case, i.e.,
$\mathcal{E}_{m}=\mathcal{E}_{m\rightarrow0}$ with $\mathcal{E}_{m\rightarrow
0}$ in (\ref{E}), where one can find the analysis of $R$ for various
parameters ranges.

For the $y$-orientation we have%
\begin{equation}
\bar{\epsilon}_{vac}^{\left(  y\right)  }=\frac{3}{R}\left(  1-\frac{2}%
{3R}\right)  \mathcal{E}_{m}~,
\end{equation}
where we used (\ref{eq:GTil}). Note that, for $\Omega=\omega_{d}$, called
zero-angular-momentum observer (ZAMO), $g_{tx}=0$ and%
\begin{equation}
\Omega=\omega_{d}\Rightarrow\bar{\epsilon}_{vac}^{\left(  y\right)  }%
=\bar{\epsilon}_{vac}^{\left(  x\right)  }=\mathcal{E}_{m}~.
\end{equation}
So (as pointed in \cite{Sorge} for the $x$-orientation), in this case the
symmetry of the spacetime is restored for all orientations. However, out of
the ZAMO configuration, the behavior of the Casimir apparatus strongly depends
on the orientation,
\begin{align}
&  \frac{\bar{\epsilon}_{vac}^{\left(  y\right)  }}{\bar{\epsilon}%
_{vac}^{\left(  x\right)  }}=1-g\left(  \delta\right)  \left[  1+2g\left(
\delta\right)  \right]  ~,\nonumber\\
&  \delta=\left\vert \Omega-\omega_{d}\right\vert \in\left[  0,\frac{r^{2}}%
{A}\sqrt{\Delta}\right)  ~,\
\end{align}
where%
\begin{equation}
g\left(  \delta\right)  =-\frac{g_{tx}^{2}}{g_{tt}g_{xx}}=\frac{\left(
A\delta\right)  ^{2}}{\Delta r^{4}-\left(  A\delta\right)  ^{2}}\in\left[
0,\infty\right)  ~.
\end{equation}
From the above expressions we see that $\bar{\epsilon}_{vac}^{\left(
y\right)  }\geq\bar{\epsilon}_{vac}^{\left(  x\right)  }$. This difference in
the energy may results in a tendency of the apparatus to assume the lowest
energy configuration, i.e., the $x$-orientation, or a torque generating a
precession around the $x$ axis.

As a special case we can considered the weak field regime, i.e., let's keep
terms up to first order in the quantities $M/r$ and $a/r$. In this case we can
write
\begin{equation}
R=1-x+O\left(  x^{2}\right)  \Rightarrow R^{-1}\simeq1+x~,
\end{equation}
which results
\begin{equation}
\frac{\bar{\epsilon}_{vac}^{\left(  y\right)  }}{\mathcal{E}_{m}}\simeq1-3x~.
\end{equation}
Which show that, in the weak field regime, the correction in the
$y$-orientation is three times greater than in the $x$-orientation. For the
example of a Casimir device resting at the equator of a spinning neutron star,
considering $M\simeq1.4M_{\odot}$, $r\simeq10^{4}%
\operatorname{m}%
$ and $\Omega\simeq190%
\operatorname{rad}%
/%
\operatorname{s}%
$, the reference \cite{Sorge} determines $x=2.3\times10^{-5}$.

As in the previous example, in the ultra-relativistic regime we have,
\begin{equation}
\Omega\rightarrow\Omega_{\pm}\Rightarrow\left\{
\begin{array}
[c]{l}%
\bar{\epsilon}_{vac}^{\left(  x\right)  }\rightarrow0\\
\bar{\epsilon}_{vac}^{\left(  y\right)  }\rightarrow\left(  \pm\right)  \infty
\end{array}
\right.  ~,
\end{equation}
with the $\left(  \pm\right)  $ sign for Dirichlet and mixed boundary
condition, respectively. Again, in the case of the $x$-orientation the Casimir
energy goes to zero, as for the Schwarzschild geometry when the orbital motion
of the cavity approaches a null geodesic orbit at $r=3M$ \cite{Sorge}. But, in
the $y$-orientation, the energy diverges for a value with a sign opposite to
$\mathcal{E}_{m}$. The Casimir force (energy) disappear at the angular
velocity%
\begin{equation}
\Omega=\Omega_{0\pm}=\omega_{d}\pm\sqrt{\frac{\Delta}{3}}\frac{r^{2}}%
{A}\Rightarrow\bar{\epsilon}_{vac}^{\left(  y\right)  }=0~. \label{o0}%
\end{equation}
%

\begin{figure}
[ptb]
\begin{center}
\includegraphics[
height=2.1473in,
width=3.3356in
]%
{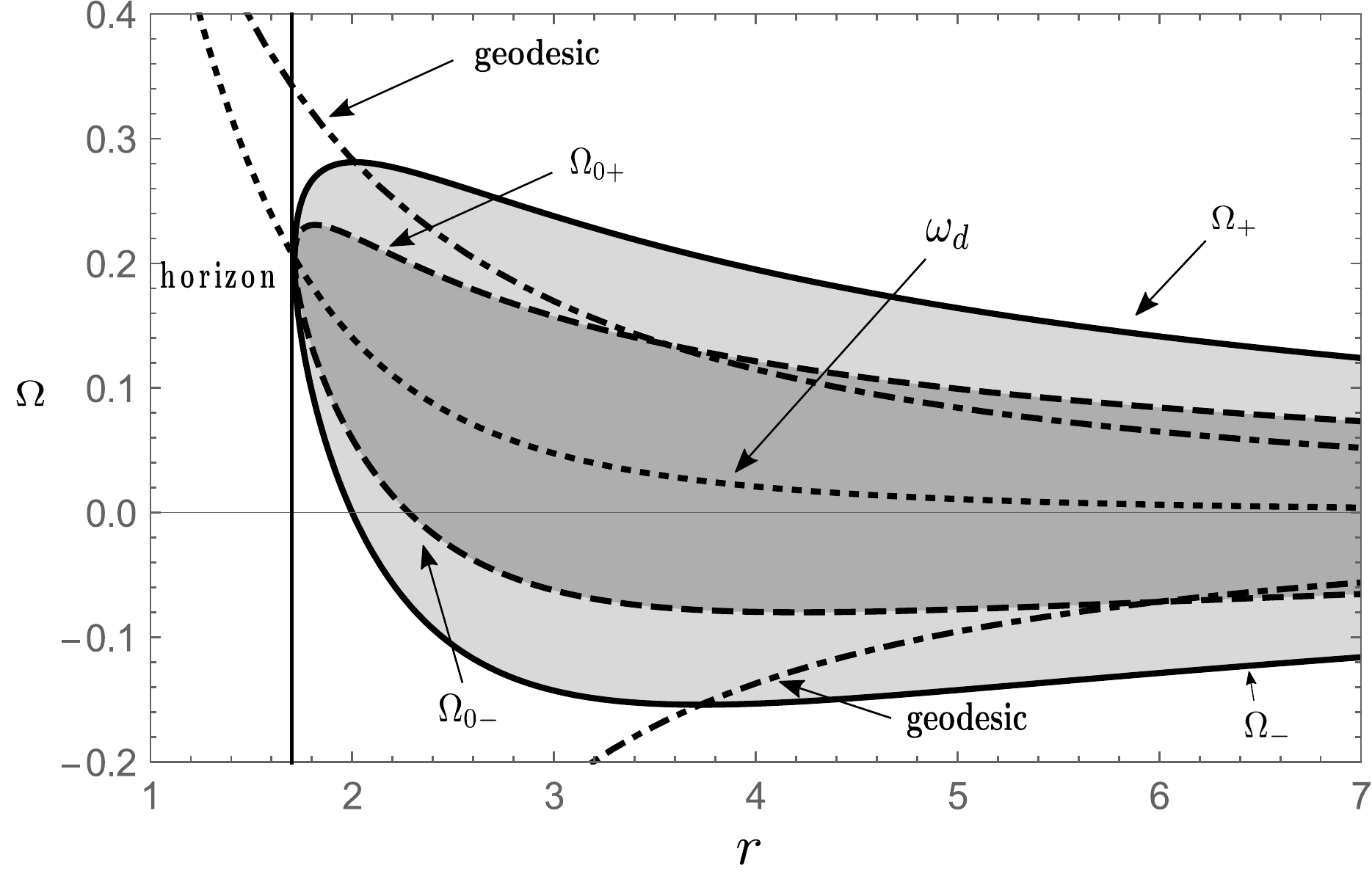}%
\caption{Plot of $\Omega$ with respect to $r$ for $M=1$ and $a=0.7$. The solid
lines $\Omega_{\pm}$ represents the ultra-relativistic cases, where the
Casimir energy disappear in the $x$-orientation. The dashed line $\Omega
_{0\pm}$ are the trajectories where the Casimir energy disappear in the
$y$-orientation. The dotted line $\omega_{d}$ the ZAMO trajectory and the
dot-dashed line the geodesic trajectories. In the $y$-orientation the energy
is negative in the dark gray region and positive in the bright gray region.
These grays regions are all the admissible trajectories.}%
\label{traj}%
\end{center}
\end{figure}

The trajectories for the velocities $\Omega_{0\pm}$ (\ref{o0})\ are shown in
Fig. \ref{traj}. In this picture $\Omega_{\pm}$ corresponded to the velocities
(\ref{pv}) when the energy tends to zero for the $x$-orientation. In the
$x$-orientation the energy is always negative. In the $y$-orientation the
energy is negative in the dark gray region and positive in the bright gray
region. The energy of both orientations coincides in the ZAMO trajectory
$\omega_{d}$. This picture shows also the geodesic trajectories. We have
geodesic trajectories where the Casimir force is attractive, repulsive, or
null, in the $y$-orientation.

\section{\label{final}Discussion}

We studied the Casimir effect in stationary axisymmetric spacetimes,
considering two orientations of the plates with respect to the drag of
spacetime. We showed that the continuous modes of the field, which contribute
to the normalization of the field and, consequently, the Casimir energy, can
change the sign of Casimir energy when the plates are perpendicular to the
direction of spacetime drag. We have explicitly showed this effect for two
examples of such metrics, one associated with a mass-energy with cylindrical
symmetry and Kerr spacetime.

Our work reproduce previous results for the massless scalar field in a
specific orientation and predict new effects for a massive scalar field. In
special, we showed that the geometry of spacetime represents a new mechanism
to change the sign of the Casimir energy, even in the absence of gravitational
coupling between the scalar and gravitational fields and mixed boundary conditions.

For the Kerr geometry we showed that the Casimir energy for the $y$%
-orientation is greater than the energy for the $x$-orientation for all
admissible circular trajectories, including the geodesics. One can expect that
this difference may result in a tendency of the apparatus to assume the lowest
energy orientation and align itself in the tangential direction with respect
to the object rotation, what can be understood as a quantum compass for
spacetime drag. Moreover, in the existence of a torque, that implies in a
precession in such direction, in a similar manner that occurs in the
Lense-Thirring effect. The effective determination of this new precession
effect requires the analyzes of the continuous variation of the orientation, a
work in progress.

Although the gravitational verification of this effects requires setups
involving the orbits around very massive and rapidly rotating objects (like
neutron stars), maybe it can be explored using some hydrodynamic analog of a
rotating black hole, as done in \cite{hydro2018} to study quasinormal modes of
such objects.

\section*{Acknowledgements}

VDC would like to thank Coordena\c{c}\~{a}o de Aperfei\c{c}oamento de Pessoal
de N\'{\i}vel Superior-Brazil (CAPES)- Finance Code 001, for financial
support, and the Universidade Estadual de Londrina for kind hospitality.

This version of the article has been accepted for publication, after
peer review (when applicable) but is not the Version of Record and does not reflect post-acceptance
improvements, or any corrections. The Version of Record is available online at:\\
http://dx.doi.org/10.1140/epjc/s10052-022-09994-4”




\end{document}